
\input amstex
\magnification 1200
\documentstyle{amsppt}
\NoBlackBoxes
\NoRunningHeads
\def\a{\alpha}
\def\l{\lambda}
\def\g{\frak g}
\def\h{\frak h}
\def\M{M}
\def\Mhat{{\widehat M}}
\def\Z{\Bbb Z}
\def\C{\Bbb C}
\def\R{\Bbb R}
\def\U{U_p\widehat{\frak{sl}}_n}
\def\eps{\varepsilon}
\def\d{\partial}
\def\i{\text{i}}
\def\ghat{\hat\frak g}
\def\hhat{\hat\frak h}
\def\H{\Cal H}
\def\What{{\widehat W}}
\def\Phat{\hat P}
\def\Pp{\hat P'}
\def\Rhat{\hat R}
\def\ahat{{\hat \alpha}}
\def\rhat{{\hat\rho}}
\def\lhat{{\hat\lambda}}
\def\gtilde{\tilde{\frak g}}

\def\Tr{\text{\rm Tr}}
\def\Id{\text{\rm Id}}

\def\v{^\vee}
\def\<{\langle}
\def\>{\rangle}

\topmatter
\title On the affine  analogue of Jack's and Macdonald's polynomials
\endtitle
\author {\rm {\bf Pavel I. Etingof, Alexander A. Kirillov, Jr.} \linebreak
\vskip .1in
Department of Mathematics\linebreak
Yale University\linebreak
New Haven, CT 06520, USA\linebreak
e-mail: etingof\@math.harvard.edu,
kirillov\@math.yale.edu}
\endauthor

\endtopmatter

\document
\subheading{ Introduction}

Jack's and Macdonald's polynomials are an important class
of symmetric functions associated to root systems.
In this paper we define and study an analogue of Jack's and
Macdonald's polynomials for affine root systems. Our approach is based
on representation theory of affine Lie algebras and quantum
affine algebras, and follows the ideas of
our recent papers \cite{EK1,EK2,EK3}.

We start with a review of the theory of Jack (Jacobi) polynomials
associated with the root system of a simple Lie algebra $\g$.
This theory was described in the papers of Heckman and Opdam
\cite{HO,H1,O1,O2}. In these papers, Jack's polynomials are defined
as a basis in the space of Weyl group invariant trigonometric
polynomials which
1) differs from the basis of orbitsums
by a triangular matrix (with respect to the standard partial ordering on
dominant integral weights) with ones on the diagonal, and
2) is an eigenbasis for a certain second order differential operator
(the Sutherland-Olshanetsky-Perelomov operator, \cite{Su,OP}).
It turns  out that these conditions determine Jack's polynomials
uniquely. Orbitsums and characters for $\g$ turn out to be special cases
of Jack's polynomials. These polynomials have a $q$-deformation, which
is called Macdonald's polynomials; they have been introduced by
I.~Macdonald in his papers \cite{M1, M2} and have been intensively
studied since that time.

We generalize the definition of Jack's polynomials
to the case of affine root systems. We assign such a polynomial
to every dominant integral weight of the affine root system.
It is done in the same way as for the usual root systems: the only
thing one has to do is
replace the Sutherland operator by its affine analogue.
This analogue is constructed in the same way as for usual root
systems,
and it turns out to be (after specialization of level) a parabolic
differential operator whose coefficients are elliptic functions.
This operator was introduced in \cite{EK3}
(for the root system $A_{n-1}$) and is closely related to the Sutherland
operator with elliptic coefficients
considered in \cite{OP}, but is more general.
Analogously to the finite-dimensional case, orbitsums and characters
(of integrable modules) for the affine Lie algebra $\hat\g$
are special cases of affine Jack's polynomials.

For orbitsums and
characters of affine Lie algebras, there is a beautiful theory
of modular invariance described in \cite{K}. We generalize this theory
to general affine Jack's polynomials. It turns out that the
finite-dimensional space spanned by the Jack's polynomials
of a given level is modular invariant with a certain weight.
Moreover, as in the character case, the representation of the modular
group in this space is (conjecturedly) unitary, with respect to a
quite nontrivial inner product
which generalizes the Macdonald inner product.
This inner product coincides with  the inner product on conformal blocks
of the Wess-Zumino-Witten conformal field theory, and its
 existence still remains a conjecture.

However, we show that
unlike the character case, the image of the corresponding
projective representation of the modular group may be infinite.

For the root system $A_{n-1}$, it is possible to give an interpretation of
Jack's and Macdonald's
polynomials in terms of representation theory of the Lie algebra
$\frak{sl}_{n}$ and quantum group $U_q(\frak{sl}_{n})$,
respectively \cite{EK1,EK2}. More specifically, Macdonald's polynomials
are interpreted as certain (renormalized) vector-valued characters (traces of
intertwiners) for quantum groups -- a notion generalizing the
usual characters. Analogously, in this paper we show that
for the root system $\hat A_{n-1}$ the
affine Jack's polynomials defined as eigenfunctions of a certain
second order
differential operator can be represented as renormalized
traces of intertwiners between certain representations
of the affine Lie algebra $\hat\g$. This proof is analogous
to the one given in \cite{EK2} for the finite-dimensional case.
Finally, we define the affine Macdonald's polynomials (i.e. q-deformed
Jack's polynomials) for the root system $\hat A_{n-1}$ to be
renormalized traces of intertwiners for quantum affine algebras,
and formulate (as a conjecture) the affine analogue of the Macdonald special
value identities from \cite{M2}.

The paper is organized as follows. In Section 1, we give basic
definitions concerning root systems. In Section 2, we define
the Sutherland operator and its eigenfunctions (Jack's polynomials)
and quote some known results about them. In Section 3, we construct
Jack's polynomials for the root system $A_{n-1}$ via representation
theory of $\frak{sl}_n$. In Section 4, we make basic definitions
concerning affine root systems. In Section 5, we define the
affine analogue of the group algebra of the weight lattice.
In Section 6, we define and study the affine Calogero-Sutherland
operator and introduce the affine Jack's polynomials. In Section 7,
we construct the affine Jack's polynomials via traces
of intertwiners for $\widehat{\frak{sl}_n}$. In Section~8,
we give a complex-analytic
description of the affine Jack's polynomials. In Section 9,
we study modular properties of the affine Jack's polynomials.
In Section~10, we give a brief introduction to the Wess-Zumino-Witten
model and formulate a conjecture on the unitarity of the
action of the modular group on affine Jack's polynomials.
 In Section~11, we define the
affine Macdonald's polynomials, and conjecture that an affine
analogue of the Macdonald special value formula is true.
Also, in this section we discuss the extension of the results
of the previous sections to non-integer values of the central charge
of the (quantum) affine algebra. Finally, Section~12 is devoted to the
discussion of some
interesting problems which still remain open.

{\bf Acknowledgements. } We would like to thank our advisor
I.~Frenkel for useful suggestions and valuable remarks.
We are grateful to H.~Garland for many interesting discussions and ideas;
his graduate course on Macdonald's polynomials given at Yale in
1993 motivated this work. We thank I.~Cherednik, I.~Grojnowski, and
D.~Kazhdan for discussions, and W.~Feit, G.~Felder, K.~Gawedzki,
G.~Margulis, and G.~D.~Mostow for information.

\subheading {\bf 1. Basic definitions: finite-dimensional case}

We let $\g$ be a finite-dimensional simple Lie algebra over $\C$, $\h$
be the Cartan subalgebra, $\text{dim } \h=r$ be the rank of $\g$,
$R\subset\h^*$ be the root system, $Q$ be the lattice in $\h^*$ spanned by the
roots, and $W$ be the Weyl group. Let us fix a decomposition of $R$
into positive and negative roots: $R=R^+\sqcup R^-$. Let
$\alpha_1,\ldots,\alpha_r\in R^+$ be the
 basis of simple roots. Then we have the
 positive cone $Q^+=\bigoplus \Z_+\alpha_i\subset Q$, and the highest
root $\theta\in R^+$ such that $\theta-\alpha\in Q^+$ for any root $\alpha$.

Let $(\cdot, \cdot)$ be a $W$-invariant bilinear symmetric form in
$\h^*$ normalized so that $(\alpha,\alpha)=2$ for long roots. This
form gives an identification $\nu\colon\h\simeq \h^*$ and thus can
also be considered as a bilinear form on $\h$.

Let $\alpha\in R$. Define the dual root $\alpha\v\in\h$ by
$\alpha\v=\frac{2\nu(\alpha)}{(\alpha,\alpha)}$,
in other words,
$\<\alpha\v,\beta\>=\frac{2(\alpha,\beta)}{(\alpha,\alpha)}$ for any
$\beta\in\h^*$ , where
$\<\cdot,\cdot\>$ is the canonical pairing between $\h$ and $\h^*$.
Then  $R\v=\{\alpha\v\}_{\alpha\in R}$ is also a root
system with the basis of simple roots given by $\alpha\v_i$. Let us
define the dual root lattice $Q\v=\bigoplus\Z\alpha\v_i$.

\proclaim{Lemma 1.1} For any $\lambda,\mu\in Q\v$, $(\lambda,\mu)\in\Z$
and $(\lambda,\lambda)\in 2\Z$. \endproclaim

Let us define the weight lattice $P=\{\lambda\in
\h^*|\<\lambda,\alpha\v\>\in\Z\text{ for any }\alpha\in R\}$ and the
cone of dominant weights $P^+=\{\lambda\in
\h^*|\<\lambda,\alpha\v_i\>\in\Z_+\}$.
Obviously, $Q\subset P$; it also follows from Lemma 1.1 that
$\nu(Q\v)\subset P$. Note that we have a partial ordering on $P$:
$\lambda\le \mu$ if $\mu-\lambda\in Q^+$. Also, it will be often used
in the future that the action of $W$ preserves $P$ and that $P^+$ is
the fundamental domain for this  action: each $W$-orbit in $P$
contains one and only one point from $P^+$.

Finally, let
$\rho=\frac{1}{2} \sum \limits_{\alpha\in R^+}\alpha$,
then $\<\rho,\alpha\v_i\>=1$ and thus $\rho\in P^+$. We
define dual Coxeter number for $\g$ by
$$h\v=(\rho,\theta)+1=\<\rho,\theta\v\>+1\tag 1.1$$

\subheading{2. Sutherland operator}
In this section we briefly summarize the known results about the
diagonalization of Sutherland operator; most results in this section
are due to Heckman and Opdam (\cite{HO; H1; O1; O2}).

Denote by $\C[P]$ the group algebra of the weight lattice; its basis
is formed by formal exponentials $e^\l,\l\in P$. We say that $f\in \C[P]$
has the highest term $e^\l$ if $f=e^\l+\sum_{\mu<\l} c_\mu e^\mu$.
We will also use the notation $f=e^\l + \text{ lower order terms}$ or
just $f=e^\l+\ldots$ in this case.
Let $\Cal R$ be the ring obtained by adjoining to $\C[P]$
the expressions of the form $(e^\alpha -1)^{-1}, \alpha\in
R$. Note that the elements of $\Cal R$ may be considered as
functions: if one replaces formal exponential $e^\lambda$ by the function
on $\h$ given by $e^\a(h)=e^{2\pi\i \<\a,h\>}$, the elements of $\Cal
R$ become functions
 on the real torus $T= \h_\R/Q\v, \h_\R=\bigoplus \R\alpha_i$
with singularities
on the hypersurfaces $e^\a (h)=1, \a\in R$. However, we will use the
formal language as far as possible.

 Let us fix some non-negative integer $k$ and
consider the following differential operator  in $\Cal R$:

$$L= L_k=\Delta - k(k-1)\sum\limits_{\alpha\in
R^+}\frac{(\a,\a)}{(e^{\alpha/2}- e^{-\alpha/2})^2},\tag 2.1$$
where $\Delta$ is Laplace's operator: $\Delta=\sum\d_{x_i}^2$,
$x_i$ being  an orthonormal basis in $\h$, and
$\d_x e^\lambda=\<x,\lambda\>e^\lambda $ with an obvious extension to
$\Cal R$. This operator for the root system $A_n$  was introduced
by Sutherland (\cite{Su}) and for an arbitrary root system by
Olshanetsky  and Perelomov
(\cite{OP}) as a Hamiltonian of an integrable quantum system. We will
call $L$  the  Sutherland operator.

Let us introduce the Weyl denominator

$$\delta=e^\rho\prod\limits_{\alpha\in\ R^+} (1-e^{-\alpha})\tag 2.2$$
and define the following version of the  Sutherland operator:

$$\M_k=\delta^{-k}(L_k-k^2(\rho,\rho))\delta^k.\tag 2.3$$

\proclaim{Lemma 2.1}{\rm (\cite{HO})}
\roster
\item $$\M_k=\Delta-k\sum\limits_{\alpha\in R^+}\frac{1+e^\alpha}
{1-e^\alpha} \d_\alpha \tag 2.4$$
(for brevity, we write $\d_\a$ instead of $\d_{\nu(\a)}$)

\item Both $L_k, \M_k$ commute with the action of the Weyl group.

\item $\M_k$ preserves the algebra $A=\C[P]^W\subset \Cal R$.
\endroster\endproclaim

Let us introduce the basis of orbitsums in $A$:

$$m_\l=\sum\limits_{\mu\in W\l}e^\mu,\qquad \l\in P^+.\tag 2.5$$

\proclaim{Lemma 2.2}
$$\M_km_\l =(\l, \l+2k\rho)m_\l+\sum\Sb \mu<\l\\ \mu\in P^+\endSb
c_{\l\mu}m_\mu\tag 2.6$$
\endproclaim

Now we can consider the eigenfunction  problem for $\M_k$. Let
us  consider the action of $\M_k$ in the finite-dimensional space
spanned by $m_\mu$ with $\mu\le \l$. Then the eigenvalue $(\l,
\l+2k\rho)$ has multiplicity one in this space due to the following
trivial but very useful fact:

\proclaim{Lemma 2.3} Let $\l,\mu\in P^+$, $\mu<\l$. Then $(\mu+\rho,
\mu+\rho)<(\l+\rho, \l+\rho)$. \endproclaim

Thus, we can give the following definition:

\proclaim{Definition} Jack's polynomials $J_\l, \l\in P^+$ are
the elements of $\C[P]^W$ defined by the following conditions:

\roster
\item $J_\l=m_\l+\sum\limits_{\mu<\l}c_{\l\mu}m_\mu$

\item $\M_kJ_\l=(\l, \l+2k\rho)J_\l$
\endroster
\endproclaim

Due to Lemma 2.3, these properties determine $J_\l$ uniquely. Note
that this definition is valid for any complex  $k$, and $J_\lambda$
are rational in $k$.

\demo{Remark} In \cite{H1}, these polynomials are called Jacobi
polynomials associated with the root system $R$; however, we prefer to
call them Jack's polynomials, since for the root system $A_n$ they are
known under this name. \enddemo

Let us introduce inner product in $A$. Let

$$\< f,g\>_0=\frac{1}{|W|}[f\bar g]_0,\tag 2.7$$
where $[\,\,]_0$ is the constant term of a polynomial, and the bar
involution is defined by $\overline{e^\l}=e^{-\l}$. More generally,
let

$$\<f,g\>_k=\<f\delta^k, g\delta^k\>_0.\tag 2.8$$

\proclaim{Lemma 2.4}
$\M_k$ is self-adjoint with respect to the inner product
$\<\cdot,\cdot\>_k$ \endproclaim
\demo{Proof} This is equivalent to saying that $L_k$ is self-adjoint
with respect to the inner product $\<\cdot,\cdot\>_0$, which is
obvious.\enddemo

\proclaim{Corollary} $\<J_\l, J_\mu\>_k=0 $ if $\l<\mu$.\endproclaim

In fact, one has a much stronger result:

\proclaim{Theorem 2.5}{\rm \cite{H1}}
$\<J_\l, J_\mu\>=0$ if $\l\ne\mu$. \endproclaim

We will prove this theorem for the root system $A_n$
 later by a different method.

Finally, let us consider the algebra $\Cal L$ of all $W$-invarinat
differential operators in $\h$ with coefficients in the ring $\Cal R$
which commute with $\M_k$. It is
proved in \cite{HO; O1; O2} that this algebra is in fact isomorphic to the
algebra of $W$-invariant polynomials in $\h$. Since it is known that
the latter one is a free polynomial algebra with generators
$p_1,\ldots,p_r, \text{ deg } p_i=d_i$, we see that $\Cal L$ is also a
free polynomial algebra generated by some differential operators
$D_i$, $D_1=\M_k$. Thus, we can formulate the eigenvalue problem:
fix a sequence
$\Lambda_1,\ldots,\Lambda_r$ and find the common eigenfunction of $D_i$ with
eigenvalues $\l_i$:
$$D_i\psi=\Lambda_i\psi, i=1\ldots r. \tag 2.9$$

In an appropriate class of functions, this system always has a
solution, and the number of solutions is equal to the order of the
Weyl group. However, if we are looking for a non-zero polynomial $W$-invariant
  solution, i.e.,
$\psi\in \C[P]^W$, then there is at most one solution for every
$\Lambda=(\Lambda_1,\ldots,\Lambda_r)$ and if it exists, it is
precisely the Jack's
polynomial $J_\l$ defined above; the corresponding eigenvalues are
$\Lambda_i=p_i(\l+k\rho), \l\in P^+$.

\subheading{3. Construction of Jack's polynomials via representation
theory}
In this section we show how one can construct the Jack's polynomials
using the representation theory of $\g$ for $\g=\frak{gl}_n$.
All the results in this section are proved in the papers (\cite{E,EK2}),
so we give them here without proofs.

For  $\l\in P^+$, let us denote by $L_\l$ the irreducible
finite-dimensional module over $\g$ with the highest weight $\l$;
also, let us denote by $V[\mu]$ the subspace of weight $\mu$ in $V$,
and let us fix a highest-weight vector  $v_\l\in V[\l]$. Let us
consider the $\g$-intertwining operators of the form
$$\Phi\colon L_\l\to L_\l\otimes U,\tag 3.1$$
where $U$ is an arbitrary module over $\g$.

Let us define the generalized character $\chi$ for such an intertwiner
by

$$\chi_\Phi=\sum\limits_{\mu\in P} e^{\mu} \Tr|_{V[\mu]}\Phi\tag 3.2$$

This is an element of $\C[P]\otimes U$; in fact, it is easy to see
that it takes values only in the zero-weight subspace $U[0]$. As in
the previous section, we can consider it as a
function on $\h$, which is
equivalent to writing

$$\chi_\Phi(h)=\Tr_V (\Phi e^{2\pi\i h}).\tag 3.3$$

One of the main results of the paper \cite{E} is the following
theorem:
\proclaim{Theorem 3.1} If $\Phi$ is an intertwiner of the form \rom{(3.1)}
then the generalized character $\chi_\Phi$ satisfies the following
equation:

$$\left(\Delta-2\sum\limits_{\a\in R^+} \frac{1}{(e^{\a/2}-e^{-\a/2})^2}
e_\a f_\a\right) (\chi\delta)= (\l+\rho, \l+\rho)\chi\delta\tag 3.4$$
\endproclaim

Another important result is the orthogonality of the generalized
characters:
\proclaim{Theorem 3.2}{\rm \cite{EK1; EK2}} If $\l,\mu\in P^+, \l\ne\mu$,
$\Phi_1\colon L_\l\to L_\l\otimes U, \, \Phi_2\colon L_\mu\to
L_\mu\otimes U$ are nonzero intertwiners, then the characters
$\chi_{\Phi_1}$ and $\chi_{\Phi_2}$ are orthogonal:
$$\sigma(\<\chi_{\Phi_1}, \chi_{\Phi_2}\>_1)=0,$$
where $\sigma: U\o U\to \C$ is the Shapovalov form.
\endproclaim

Now, let us be more specific. From now till the end of this section,
we only consider the Lie algebra $\frak s\frak l_n$, i.e., the root system
$A_{n-1}$. Let us fix a positive  integer $k$ (later it will be the
same $k$ we considered in Section 2) and take $U=S^{(k-1)n}\C^n$. Then
$U[0]$ is one-dimensional, and for every $\a\in R^+$,
$e_\a f_\a|_{U[0]}=k(k-1)$.

\proclaim{Lemma 3.3} Let $\mu \in P^+$. A non-zero intertwiner

$$\Phi\colon L_\mu \to L_\mu\otimes U$$
exists iff $\mu= (k-1)\rho+ \l, \l\in P^+$; if it exists, it is unique up to a
scalar. We will denote such an intertwiner by $\Phi_\l$.
\endproclaim
Proof of this lemma is a standard exercise, which we leave to
the reader.

Since $U[0]\simeq\C$, we can consider $\chi_\Phi$ as a scalar-valued
function;  we choose this identification in such a way that
$$\chi_\l=\chi_{\Phi_\l} = e^{\l+(k-1)\rho}+\text{ lower order terms}.$$

Now we quote  two  results from \cite{EK2}:

\proclaim{Theorem 3.4}
$$\chi_0=\delta^{k-1}.\tag 3.5$$
\endproclaim

\proclaim{Theorem 3.5} $\chi_\l$ is divisible by $\chi_0$, and the
ratio is the Jack's polynomial:
$$\frac{\chi_\l}{\chi_0} =J_\l.$$
\endproclaim

\demo{Proof} We briefly outline the proof, since it is very
instructive. First, we prove by induction in $\l$ that $\chi_\l$ is
divisible by $\chi_0$ and the ratio is a symmetric polynomial with the
highest term $e^\l$. Next, it follows from Theorems 3.1 and 3.4 that

$$\left(\Delta-2k(k-1)\sum\limits_{\a\in R^+} \frac{1}{(e^{\a/2}-e^{-\a/2})^2}
\right) \left(\frac{\chi_\l}{\chi_0}\delta^k\right)= (\l+k\rho,
\l+k\rho)\frac{\chi_\l}{\chi_0}\delta^k.$$

Comparing this with the formulas for operators $L_k$ and $\M_k$ in
Section~1, we see that $\chi_\l/\chi_0$ is the Jack's polynomial $J_\l$.
\qed\enddemo

\proclaim{Corollary} $\<J_\l, J_\mu\>_k=0$ if $\l\ne \mu$.\endproclaim
\demo {Proof} This  follows immediately from Theorems 3.2, 3.4 and
3.5.\enddemo

\subheading{ 4. Basic definitions: affine case}
Here  we review  the notations and facts about affine
Lie algebras and root systems. All of them can be found in \cite{K}.
 As a rule, we will use hat (\ $\hat{}$ ) in the notations of
affine analogues of finite-dimensional objects.

Let $\ghat$ be the affine Lie algebra corresponding to $\g$:

$$\ghat=\g\otimes \C[t,t^{-1}]\oplus\C c \oplus\C d,$$
with the commutation rule given by

$$\gathered
[x\otimes t^n,y\otimes t^m]=[x,y]\otimes t^{m+n}+n\delta_{m,-n}
(x,y)c\\
c \text{ is central}\\
[d,x\otimes t^n]=nx\otimes t^n\endgathered\tag 4.1$$

Sometimes we will use a smaller algebra $\gtilde=\g\otimes
\C[t,t^{-1}]\oplus\C c $.

Similarly to the finite-dimensional case, we define Cartan subalgebra
$\hhat=\h\oplus\C c\oplus \C d$, $\hhat^*=\hhat^*\oplus\C
\delta\oplus\C\eps$, where $\<\eps, \h\oplus\C d\>=\<\delta, \h\oplus
\C c\>=0$, $\<\delta,d\>=1,\<\eps, c\>=1$.
It will be convenient to consider affine hyperplanes
$\hhat^*_K=\h^*\oplus\C \delta +K\eps, K\in \C$; we will refer to the
elements of $\hhat^*_K$ as having level $K$.

Again, we have a bilinear non-degenerate symmetric form $(\cdot,
\cdot)$ on $\hhat^*$ which coincides with previously defined  on
$\h^*$  and $(\eps, \delta)=1, (\eps,\h^*)=(\delta,
\h^*)=(\eps,\eps)=(\delta,\delta)=0$.
This gives an identification $\nu\colon \hhat\simeq \hhat^*$,
and a bilinear form on $\hhat$ such that $(c,d)=1$.

We define the root system $\Rhat=\{\hat\alpha=\alpha+n\delta|\alpha\in
R,n\in\Z \text{ or }\alpha=0, n\in\Z\setminus\{0\}\}$. Again, we have
the notion of positive roots: $\Rhat^+=\{\hat\alpha=\alpha+n\delta\in
\Rhat|n>0 \text{ or } n=0, \alpha\in R^+\}$ and the basis of simple
roots $\alpha_0=-\theta+\delta,\alpha_1,\ldots,\alpha_r$.

Now, let us define the affine Weyl group $\What$ as the group of
transformations of $\hhat^*$ generated by the reflections with respect
to $\alpha_i, i=0\ldots r$. This group  preserves the bilinear form;
also, it preserves each of the affine hyperplanes $\hhat^*_K$. We have
notion of sign of an element of $\What$: $\eps(w)=(-1)^l$ if $w$ is a
product of $l$ reflections.

\proclaim{Theorem 4.1}{\rm (see \cite{K})}
 $\What\simeq W\ltimes Q\v$, where the action of $W$ is the same as in
the classical case, and the action of $Q\v$ in $\hhat^*_K$ is given by
$$\alpha\v\colon \lhat \mapsto \lhat+ K\nu(\alpha\v)-\bigl(
\<\lhat,\alpha\v\> +\frac{1}{2} K(\alpha\v,\alpha\v)\bigr)\delta\tag
4.2$$\endproclaim

Now we can define the root lattice $\Phat=P\oplus\Z\delta\oplus
\Z\eps\subset \hhat^*$ and the cone of dominant weights $\Phat^+= \{
\lhat\in\hhat^*| \<\lhat,\alpha\v_i\>\in\Z_+, i=0,\ldots,r\}$. We will
also use the notation $\Phat^+_K=\Phat^+\cap
\hhat^*_K=  \{\lambda+n\delta+K\eps| \lambda\in P^+,
\<\lambda,\theta\v\>\le K\}$ for $K\in \Z_+$. Note
the cone of dominant weights is invariant with respect to the translations
along $\delta$ direction, but if one factors this out then there is
only a finite number of dominant weights for every level $K$. Abusing
the notations, we will write $P^+_K= \{\lambda\in P^+|
(\lambda,\theta)\le K\}$.  Also, we introduce the following affine
analogue of $\rho$: $$\hat\rho=\rho+h\v\eps$$ then $\<\hat\rho,
\alpha_i\v\>=1, i=0,\ldots,r$ and thus $\hat\rho\in
\Phat^+$.

\proclaim{Lemma 4.2} 1. $\What$ preserves each $\Phat_K=\Phat\cap\hhat^*_K$

2. $\Phat^+_K$ is a fundamental domain for the action of $\What$ in
$\Phat_K$ for $K>0$.\endproclaim

\subheading{5. The group algebra of the weight lattice}
In this section we define our basic object of study -- the algebra of
$\What$-invariants of the (suitably completed) group algebra of
$\Phat$ and study its elementary properties, following the paper of
Looijenga (\cite{Lo}).

 Let us consider the group algebra $\C[\Phat]$, i.e. the algebra
spanned by the formal exponentials $e^{\lhat}, \lhat\in \Phat$. It is
naturally $\Z$-graded: $\C[\Phat]=\bigoplus\limits_{K\in\Z}
\C[\Phat_K]$. Consider the following completion:

$$\overline{\C[\Phat_K]}=\{\sum\limits_{n=1}^\infty
a_ne^{\lhat_n}|\lim_{n\to\infty}(\lhat_n,\hat\rho)= -\infty\}\tag 5.1$$

Then $\overline {\C[\Phat]}=\bigoplus\limits_K\overline{\C[\Phat_K]}$
is again a $\Z$-graded algebra. (This completion is chosen so to
include the characters of Verma modules over $\ghat$.)

However, we will use a smaller algebra $A=\bigcap\limits_{w\in
\What} w\biggl(\overline{\C[\Phat]}\biggr)$, which is a natural
analogue of the group algebra $\C[P]$ for finite-dimensional case. In
particular, we have a natural action of $\What$ in $A$. It is also
$\Z$-graded: $A_K=\bigcap\limits_{w\in
\What} w\biggl(\overline{\C[\Phat_K]}\biggr)$

Now, let us consider the algebra of $\What$-invariants
$A^{\What}\subset A$. Abusing the language, we will call elements of
$A^\What$ invariant polynomials.

\demo{Example 1} For any $\lhat\in \Phat_K, K\ge 0$ the orbitsum
$m_{\lhat}=\sum\limits_{\hat\mu\in \What\lhat}e^{\hat\mu}$ belongs to
$A^{\What}_K$.\enddemo

\demo{Example 2} If $\lhat$ is a dominant weight then the irreducible
highest-weight module $L_{\lhat}$ with highest weight $\lhat$ is
integrable, and its character belongs to $A^{\What}_K$.\enddemo

Obviously, $A^{\What}$ is $\Z$-graded. Moreover, the following is
well-known:

\proclaim{Lemma 5.1} $A^{\What}_K=0$ for $K<0$, and
$A^{\What}_0=\left\{\sum\limits_{n\le n_0}a_ne^{n\delta} ,
a_n\in\C\right\}$ \endproclaim

\proclaim{Theorem 5.2}{\rm (cf. \cite{Lo})} For every $K\in\Z_+$, the
orbitsums $m_{\lambda+K\eps}, \lambda\in P^+_K$ form a basis of $A^{\What}_K$
over
the field $A^{\What}_0$.\endproclaim

This theorem follows from the fact that $\Phat^+_K$ is a fundamental
domain for the action of $\What$ in $\Phat_K$.

It will be convenient to introduce formal variable $q=e^{-\delta}$;
then every element of $A^{\What}$ can be written as a formal Laurent series in
$q$ with coefficients from $\C[P]$ (this relies on $\What$
invariance). In particular, in these notations $A^{\What}_0\simeq \C((q))$.

\subheading{ 6. The affine Calogero-Sutherland operator}
In this section we give the definition of the affine analogue of
Sutherland operator, which we will call the affine Calogero-Sutherland
operator. As before, we fix some positive integer $k$.

First of all, let us define the analogue of the ring $\Cal
R=\C[P](1-e^\a)^{-1}$, defined in Section~2.  Consider the algebra
$\C[\Phat](1-e^\a)^{-1}$, obtained by adjoining to $\C[\Phat]$ the
inverses of $(1-e^\a)$ for $\a\in \Rhat$ (no completion so far).
Then we have a morphism
$$\tau\colon \C[\Phat](1-e^\a)^{-1}\to \overline{\C[\Phat]},$$
given by expanding $(1-e^{-\a})^{-1}=1+e^{-\a}+e^{-2\a}+\ldots$ for
$\alpha\in \Rhat^+$. Note that the image is not  in $A$.

Similarly, for every $w\in\What$ we have

$$\tau_w\colon \C[\Phat](1-e^\a)^{-1}\to
 w\biggl( \overline{\C[\Phat]}\biggr),$$
given by expanding $(1-e^{-\a})^{-1}=1+e^{-\a}+\ldots$ for $\a\in w\Rhat^+$.

Define $\hat\Cal R=\{\sum a_n|a_n\in \C[\Phat](1-e^\a)^{-1}, \sum
\tau_w(a_n) \text{ converges in } w\biggl(\overline{\C[\Phat]}\biggr)
\text{ for every }w\in \What\}$. This is the right analogue of the
ring $\Cal R$ introduced in Section~2; for example, $\sum_{\ahat\in
\Rhat^+} \frac{1}{1-e^\ahat}\in \hat\Cal R$. Note that there is a natural
action of the Weyl group $\What$ in $\hat\Cal R$; also note that this
algebra has a natural $\Z$-grading given by level.

\definition{Definition}
The Calogero-Sutherland operator for (affine) root system $\Rhat$ is
the differential operator which acts in $\hat\Cal R_K$ by
the following  formula (all the notations as before):

$$\hat L_k= \Delta - 2K q\frac{\d }{\d q}-k(k-1)\sum\Sb \a \in R^+\\n\in
\Z\endSb
\frac{q^ne^{\a}}{(1-q^ne^{\a})^2}(\a,\a).\tag 6.1$$
\enddefinition

\remark{Remark} Note that for $K=0$ this operator coincides (up to a
constant) with the elliptic Calogero-Sutherland operator
(see \cite{OP}); compare with formula (8.5) below.
In more general situation, operetor (6.1) was introduced by Bernard
(\cite{B}).
\endremark

Introducing the Laplace's operator $\hat\Delta$ on $\hhat$: $\hat\Delta=
\Delta+2\d _c\d_d$, we can rewrite (6.1) as follows:

$$\hat L=\hat\Delta -k(k-1)
\sum\limits_{\hat\a\in\Rhat^+}\frac{1}{(e^{\hat\a/2}-e^{-\hat\a/2})^2}
(\hat\a,\hat\a),\tag 6.2$$
making $\hat L$ an absolute analogue of (2.1). Note that $\hat L$ obviously
commutes with the action of $\What$ in $\hat\Cal R$.

Similarly to Section 2, define

$$\hat\delta=
e^{\rhat}\prod\limits_{\hat\a\in\Rhat^+}(1-e^{-\hat\a}).\tag 6.3$$

This is an element of $A$. It is known (see \cite{Lo, K})
that $\hat\delta$ is
$\What$-antiinvariant; moreover, every $\What$-antiinvariant element
of $A$ has the form $f\delta, f\in A^{\What}$.

Define
$$\Mhat_k=\hat\delta^{-k}\circ(\hat L_k-k^2(\rhat,\rhat))
\circ\hat\delta^k.\tag 6.4$$

\proclaim{Theorem 6.1}\roster
\item $\Mhat_k$ is a well defined operator in $\hat\Cal R$

\item
$$\Mhat_k=\hat\Delta-2k\sum\limits_{\hat\a\in\Rhat^+}
\frac{1}{1-e^{\hat\a}}\d_{\hat\a} +2k \d_{\rhat}.\tag 6.5$$

\item $\Mhat_k$ commutes with the action of $\What$.
\endroster
\endproclaim

\demo{Proof} It will be convenient to use vector fields in $\hhat$, i.e.
the elements of $\hat\Cal R\otimes \hhat$. If $f=\sum f_i\lambda_i,
f_i\in \hat\Cal R, \lambda_i\in \hhat$ then we will denote by
$\d_f=\sum f_i\d_{\lambda_i}$ the corresponding differential operator.
Then we have the following obvious formula:

$$\hat\delta^{-1}\circ \hat \Delta \circ \hat \delta =
\hat\Delta + 2\d_v + (\hat\delta^{-1}\hat\Delta(\hat\delta)),\tag 6.6$$
where

$$v=\hat\delta^{-1}\text{grad }\hat\delta=
\rhat+\sum_{\hat\a\in\Rhat^+}\frac{e^{-\hat\a}}{1-e^{-\hat\a}} \hat\a.$$
Note that $\What$-antiinvariance of $\hat\delta$ implies
$\What$-invariance of $v$. Since $\hat\Delta
(\hat\delta)=(\rhat,\rhat)\hat\delta$, which follows from the
denominator identity for affine root systems,  this proves that
$\hat\delta^{-1}\hat\Delta \hat\delta$ is a well defined operator in
$\hat\Cal R$.

It is easy to prove by induction that

$$\hat\delta^{-k}\circ \hat\Delta\circ \hat\delta^k =
\hat\Delta + k(\rhat, \rhat) + 2k\d_v + k(k-1) \hat\delta^{-1}
(\d_v\hat\delta).$$

Obviously, $\hat\delta^{-1}(\d_v\hat\delta)=(v,v)$, where $(\,,\,)$ is
the inner product of the vector fields, i.e., the inner product on
$\hhat$ extended by $\hat\Cal R$-linearity to $\hat\Cal R\otimes
\hhat$ (it has nothing to do with the inner product on  polynomials!).

\proclaim{Lemma}
$$(v,v)= (\rhat, \rhat)+\sum\limits_{\hat\a\in \Rhat^+}
(\hat\a,\hat\a)\frac{1}{(e^{\hat\a/2} -e^{-\hat\a/2})^2}.$$
\endproclaim

\demo{Proof} Let us consider $X=(v,v)-(\rhat,\rhat)-\sum_{\hat\a\in \Rhat^+}
(\hat\a,\hat\a)\frac{1}{(e^{\hat\a/2} -e^{-\hat\a/2})^2}\in\hat\Cal
R$. Obviously, it is $\What$-invariant. Also, from the explicit
expression for $v$ it follows that $X$ has only simple poles.
Consider $X\hat\delta$. It is an element of $\hat\Cal R$
with no poles, thus it is an element of $A$. Also, it is
antiinvariant. Thus, $X\in A$. Obviously, $X\in A_0$; since $X$ is
$\What$-invariant, Lemma~5.1 implies that $X\in \C((q))$.

To complete the calculation, let us write the explicit expression for
$X$; then, let us expand it in a series in $e^{-\hat\a},\hat\a\in
\Rhat^+$ (i.e., apply the map $\tau$ defined above) and keep only
the terms of the form $e^{n\delta}$ in the expansion. This gives:

$$X=2\biggl(rh\v - \sum\limits_{\a\in
R^+}(\a,\a)\biggr)\sum\limits_{n=1}^\infty \frac{nq^n}{1-q^n} $$

On the other hand, it is known
that for any Lie algebra $\g$, $\sum_{\a\in R^+}(\a,\a)= rh\v$. The
simplest way to prove it is to consider the action of the Casimir
element $\Omega\in U\g$ in the adjoint representation. On one hand,
$\Omega|_{\g} =2h\v\Id _{\g}$, and thus $\Tr|_{\h}\Omega= 2rh\v$. On
the other hand, it is easy to deduce from the formula
$\Omega=\sum_{\a\in R^+}e_\a f_\a +f_\a e_\a +\sum x_i^2$ that
$\Tr_{\h} \Omega = 2\sum_{\a\in R^+} (\a,\a)$.
 Thus,  $X=0$. \enddemo

This lemma together with previous results immediately implies statements
1 and 2 of the theorem. Statement 3 follows from $\What$-invariance of
the operator $L_k$.
\qed\enddemo

Note that Theorem 6.1 is a complete analogue of Lemma 2.1.
\remark{Remark 1} This technique is borrowed from \cite{Ma}.
\endremark

\remark{Remark 2} In the simply-laced case the identity $\sum(\a,\a)=rh\v$
becomes $\text{dim }\g=r(h+1)$, where $h$ is the Coxeter number for
$\g$. This latter identity is well known and has
a beautiful interpretation in terms of the Coxeter automorphism, due
to Kostant. \endremark

\proclaim{Theorem 6.2}
$\Mhat$ preserves the algebra of $\What$- invariant polynomials: $\Mhat
A^{\What}_K\subset A^{\What}_K$. Moreover, its action is triangular:

$$\Mhat m_\lhat=(\lhat, \lhat+2k\rhat)m_\lhat+\sum\Sb \hat\mu<\lhat\\
\hat\mu\in \hat P^+_K \endSb c_{\lhat\hat\mu}m_{\hat\mu}\tag 6.7$$
\endproclaim

\demo{Proof} Proof is based on the following lemma:

\proclaim {Lemma} Let $f\in A^{\What}, f=e^\lhat +\text{ lower
terms}, \hat\a\in \Rhat^+$. Then
$\frac{1}{1-e^{\hat\a}}\d_{\hat\a} f$ is
a well-defined element of $A$ with highest term
$-(\lhat,\hat\a)e^{\lhat-\hat\a}$. \endproclaim

\demo{Proof} The proof is based on the fact that due to
$\What$-invariance, $f$ contains terms $e^{\hat\mu}$ and $e^{\hat\mu-
\<\hat\mu,\hat\a\v\>\hat\a}$ with equal coefficients, and on explicit
calculation. \enddemo

Now the statement of the theorem follows from Theorem 6.1.
\qed\enddemo

Our main objective will be the study of the eigenfunctions of action
of this operator in $A^{\What}$, which we will call affine Jack's
polynomials.  More precisely, let us consider the action of $\Mhat$
in the linear space spanned by $m_{\hat\mu}$ with $\hat\mu\le\lhat$. This
space is not finite-dimensional; however, one can still check that
$\Mhat$ has a unique eigenvector with eigenvalue $(\lhat,\lhat+2k\rhat)$ in
this space (this is based on the affine analogue of Lemma 2.3).
Thus, we adopt the following definition:

\proclaim{Definition} Affine Jack's polynomials $\hat J_\lhat, \lhat\in
\Phat^+$  are
the elements of $A^{\What}$ defined by the following conditions:

\roster
\item
$\hat J_\lhat=m_\lhat+\sum\limits_{\hat\mu<\lhat}c_{\lhat\hat\mu}m_{\hat\mu} $

\item $\Mhat_k \hat J_\lhat=(\lhat, \lhat+2k\rhat)\hat J_\lhat$
\endroster
\endproclaim

As was said above, these conditions determine $J_\lhat$ uniquely.
Note that if $\hat\mu=\lhat+n\delta$ then $J_{\hat\mu}=q^{-n}J_\lhat$.
Thus, it suffices to consider only the polynomials $J_\lhat$ for
$\lhat=\l+K \eps, \l\in P^+_K$.

\subheading{7. Intertwiners and traces for affine Lie algebras}

In this section we introduce the tools we will use later to construct
affine Jack's polynomials.

First, we define the notion of evaluation representation. Let $V$ be a
finite-dimensional module over $\g$ and $z$ be a non-zero complex
number. Then we can construct an evalution representation of $\gtilde$
(not $\ghat$!) in $V$ by

$$\pi_{V(z)}(a\otimes t^n)=z^n\pi_{V}(a), \pi_{V(z)}(c)=0.$$

Note that $V(z)$ has no $\Phat$-gradation but has a natural
$P$-gradation.

We will be interested in intertwining operators

$$\Phi\colon L_{\lhat}\to \widehat{L_{\lhat}}\otimes V(z),\tag 7.1$$
where $\widehat{L_\lhat}$ is the completion of the integrable highest-weight
module with respect to the $d$-grading.
To prove the existence of such intertwiners, we use the following
well-known result (see, for example, the arguments in \cite{TK}, which
work for general Lie algebra in the same way as for $\frak s \frak l_2$):

\proclaim{Lemma 7.1}
The mapping $\Phi\mapsto \<v_{\lhat}, \Phi v_{\lhat}\>$ establishes
one-to-one correspondence between the space of all intertwiners of the
form \rom{(7.1)} and the subspace in $V[0]$ formed by the vectors $v$ such
that
$x v=0$ for every  $x\in U\ghat^-$ such that
$xv_\lhat=0$ in $L_\lhat$.
\endproclaim

Suppose that $\Phi$ is a non-zero intertwiner of the form (7.1). Then
we define  the ``generalized character'' $\chi_\Phi$ by

$$\chi_\Phi=\sum\limits_{\hat\mu\in \Phat}e^{\hat\mu}
\Tr|_{L_{\lhat}[\hat\mu]}\Phi .\tag 7.2$$

This is an element of $\overline{\C[\Phat]}\otimes V[0]$ with the
highest term $\<v_\lhat, \Phi v_\lhat\>e^\lhat$, and it is
independent of  $z$.  Moreover, let
us assume that $\lhat$ is a dominant weight. Then $\chi_\Phi\in A$.
Unless stated otherwise, in
this chapter we always assume that $\lhat$ is dominant.

\proclaim{Theorem 7.2}{\rm (see \cite{EK3})} If $\Phi$ is an intertwiner of
the form (7.1), $\lhat=\l+a\delta+K\eps$ then the generalized character
$\chi_\Phi$ satisfies the following equation:

$$\biggr(\Delta-2(K+h\v)q\frac{\d}{\d q} -2 \sum\Sb \a\in R^+\\n\in\Z\endSb
\frac{q^ne^\a}{(1-q^ne^\a)^2}e_\a
f_\a\biggl)(\chi_\Phi\hat\delta)=(\lhat+\rhat,
\lhat+\rhat)\chi_\Phi\hat\delta\tag 7.3$$
\endproclaim

As before, we restrict ourselves to the case $\g=\frak{sl}_n$, and
take $V=S^{(k-1)n} \C^n$. Then $V[0]\simeq\C$, and
$e_\a f_\a|_{V[0]}=k(k-1)$.

\proclaim{Proposition 7.3} A non-zero intertwiner
$$\Phi\colon L_{\hat\mu}\to \widehat{L_{\hat\mu}}\otimes V(z)$$
exists iff $\hat\mu=(k-1)\rhat +\lhat, \lhat\in \Phat^+$; if it
exists, it is unique up to a scalar. We will denote such an
intertwiner by $\Phi_\lhat$.
\endproclaim

\demo{Proof} The proof is based on Lemma 7.1.\enddemo

Let us consider the traces $\chi_\lhat=\chi_{\Phi_\lhat}$. They take
values in $V[0]$, which is one-dimensional, and thus can be considered
as scalar-valued; we choose this identification so that
$$\chi_\lhat= e^{\lhat+(k-1)\rhat} +\text{ lower terms}$$

\proclaim{Proposition 7.4} For every real $\hat\a\in \Rhat^+$,
$\chi_\lhat$ is divisible by $(1-e^{-\hat\a})^{k-1}$ (divisibility is
to be understood in the algebra $A$).
\endproclaim
\demo{Proof}  The proof is absolutely similar to the
finite-dimensional case (cf. \cite{EK2}), and is based on consideration
of the traces of the form
$$\chi_\lhat^F=\sum\limits_{\hat\mu\in \Phat}e^{\hat\mu}
\Tr|_{L_{\lhat}[\hat\mu]}(\Phi_\lhat F) , \tag 7.4$$
where $F$ is an arbitrary element of $U\ghat$.  Let us take $F=f_{\hat\a}
^{k-1}$. Then, using the intertwinig property of $\Phi_\lhat$ and the
identity $\Delta (f_\ahat)=f_\ahat\otimes 1+ 1\otimes f_\ahat$, we can
prove by induction that

$$\chi_\lhat^F=\frac{f_\ahat
^{k-1}\chi_\lhat}{(1-e^{-\ahat})^{k-1}}.$$

Since $V[0], V[-(k-1)\a]$ are one-dimensional, we can identify both of
them with $\C$; then
$f_\ahat^{k-1}\colon V[0]\to V[-(k-1)\a]$ becomes a non-zero constant.
On the other hand, it is easy to see that $\chi_\lhat^F\in A$, which
proves the proposition.\qed\enddemo

\proclaim{Theorem 7.5}
$$\chi_0=\hat\delta^{k-1}.$$
\endproclaim
\demo{Proof} Let us consider the ratio $f=\chi_0/\hat\delta^{k-1}$. It
follows from Proposition 7.4 that
$f\in A$. It has level zero and highest term $1$. Moreover,
similar arguments show that if we twist the order on $\Phat$ by the
action of the Weyl group: $\lhat\ge_w\hat\mu$ if $\lhat-\hat\mu\in
w(Q^+), w\in \What$ then highest term of $f$ with respect to any such
twisted ordering is still $1$. This is only possible if $f\in \C((q))$.
To complete the proof,  we have to use the differential equation
for the characters. Indeed, Theorem 7.2 implies that $\chi_\lhat$
satisfies the following equation:

$$\hat L (\chi_\lhat \hat\delta)
=(\lhat+k\rhat,\lhat+k\rhat)(\chi_\lhat\hat\delta).$$
Substituting in this equation $\chi_0=f(q)\hat\delta^{k-1}$, we see that
$f$ satisfies $\Mhat f=0$. Using   formula (6.5)  for
$\Mhat$ we get  $2kh\v q\frac{\d}{\d q}f=0$,
which  is possible only if $f$ is a constant. Comparing highest terms of
$\chi_0$ and $\hat\delta^{k-1}$, we get the statement of the theorem.\qed
\enddemo

\proclaim{Theorem 7.6}

$$\frac{\chi_\lhat}{\chi_0}=\hat J_\lhat.$$

\endproclaim
\demo{Proof}
Let us first prove that $\chi_\lhat/\chi_0\in A^{\What}$. Consider the module
$L=L_\lhat\otimes L_{(k-1)\rhat}$. This  module
unitary (since both factors are unitary); thus, it is
completely reducible and can be decomposed in a direct sum of the
modules $L_{\hat\mu}$:

$$L=L_{(k-1)\rhat+\lhat}+\sum\Sb \hat\mu\in\Phat^+\\
\hat\mu<\lhat+(k-1)\rhat\endSb N_{\hat\mu}L_{\hat\mu}$$

This sum is, of course, infinite; however, all the multiplicities are
finite. In particular, this implies that the character of this module
belongs to the algebra $A^{\What}$ (that is, $\What$-invariants of completed
group algebra of $\Phat$), so in a certain sense this sum converges.

Let us construct an intertwiner $\Psi\colon L\to L\otimes V(z)$ as
$\Psi= \Id_{L_\lhat}\otimes \Phi_0$. Consider the corresponding
trace  $\chi_\Psi$. Then it follows from the decomposition of $L$
that

$$\chi_\Psi=\chi_\lhat+\sum\Sb \hat\mu\in\Phat^+\\ \hat\mu<\lhat\endSb
a_{\lhat\hat\mu}\chi_{\hat\mu}.$$

On the other hand, $\chi_\Psi=\chi_0\text {Ch }L_\lhat$, where
$\text{Ch } L_\lhat$ is the ordinary character of the module $L_\lhat$. Thus,
dividing both sides by $\chi_0$ we get

$$\text{ Ch }L_\lhat=\sum\limits_{ \hat\mu\in\Phat^+} a_{\lhat\hat\mu}
\frac{\chi_{\hat\mu}}{\chi_0},$$
where $a_{\lhat\hat\mu}\ne 0$ only if $\hat\mu\le \lhat$, and
$a_{\lhat\lhat}=1$. It is a trivial exercise in linear algebra to
check that in this case the matrix $(a_{\lhat\hat\mu})$ has an
inverse: one can write
$$\frac{\chi_\lhat}{\chi_0}=\sum\limits_{\hat\mu\in \Phat^+}
b_{\lhat\hat\mu} \text{ Ch }L_{\hat\mu},$$
and the coefficients $b_{\lhat\hat\mu}$ satisfy the same conditions as
$a_{\lhat\hat\mu}$. Thus, $\chi_{\lhat}/\chi_0\in A^{\What}$ and has
highest term $e^\lhat$.

We have proved that $\chi_\lhat/\chi_0$ satisfies the first
condition in the definition of the Jack's polynomials. Now,
Theorem~7.2 implies that $\hat L (\chi_\lhat
\hat\delta)=(\lhat+k\rhat,\lhat+k\rhat)
(\chi_\lhat\hat\delta)$. Due to Theorem~7.5, this means that

$$\hat L \biggl(\frac{\chi_\lhat}{\chi_0}\hat\delta^k\biggr)
=(\lhat+k\rhat, \lhat+k\rhat)
\biggl(\frac{\chi_\lhat}{\chi_0}\hat\delta^k\biggr),$$
which is precisely the definition of Jack's polynomials. \qed\enddemo

\subheading{ 8. Normalized characters and their functional
interpretation}

This section is of preparatory nature; its results will be used in the
next section where we study the modular properties of the Jack's
polynomials.

First of all, to make our functions modular invariant we need to
introduce some factors of the form $q^t, t\in \Bbb Q$. Thus, we need
to consider slightly more general setting than in
the previous section. Namely, instead of the weight lattice $\Phat$ we
consider a bigger abelian group $\Pp =P+\C\delta +\Z\eps$.
Also, we can consider
the algebra $A'=\{\sum_{i=1}^N q^{a_i} f_i, a_i\in \C, f_i\in A\}$, and
the subalgebra of $\What$-invariants in $A'$ in a manner quite similar to the
one of the previous section. All the results of Section~5 hold with
obvious changes.

Define the normalized analogues of $\rhat$ and $\hat\delta$ as
follows:
$\rhat'=\rhat-\frac{(\rho,\rho)}{2h\v}\delta=
\rho-\frac{(\rho,\rho)}{2h\v}\delta+h\v\eps$,
$\hat\delta'=e^{\rhat'} \prod_{\ahat\in\Rhat^+}
(1-e^{-\ahat})$. This renormalization is chosen so that $\hat\Delta
\hat\delta'=0$; another reason for this renormalization is that so
defined $\hat\delta'$ possesses nice modular properties (see below).

Now, let us define the renormalized operator

$$\Mhat'=\hat\delta^{\prime -k}\hat L \hat\delta^{\prime k}=
\Mhat -\frac{Kk(\rho,\rho)}{h\v}.\tag 8.1$$

Finally, for $K\in Z_+, \lambda\in P^+_K$, (i.e., $(\lambda,\theta)\le
K$) consider

$$\lhat=\lambda+K\eps + \delta\biggl(\frac{(\rho,\rho)}{2h\v}
-\frac{(\lambda+k\rho,\lambda+k\rho)}{2(K+kh\v)}\biggr).\tag 8.2$$

Now we can consider the generalized characters $\chi_\lhat$, defined in
Section~7, and introduce the normalized Jack's polynomials:

\definition{Definition} If $K\in\Z_+, \lambda\in P^+_K$ then the
Jack's polynomial $J_{\lambda, K}$ is given by

$$J_{\lambda, K}=\frac{\chi_\lhat}{\hat\delta^{\prime (k-1)}},$$
where $\lhat$ is given by (8.2).\enddefinition

The results of the previous section imply that $J_{\lambda, K}$ are
invariant polynomials: $J_{\lambda, K}\in A^{\prime\What}$, and that highest
term of $J_{\lambda, K}$ is
$e^{\lambda+K\eps+\delta\left(\frac{k(\rho,\rho)}{2h\v}
		-\frac{(\lambda+k\rho,
\lambda+k\rho)}{2(K+kh\v)}\right)}$.

Note that for $k=1$ they are precisely the (usual) characters of
integrable highest-weight modules, and the normalization coincides
with that in \cite{K, Chapter 13}.

Moreover, it follows from Theorem~7.6 that the normalized Jack's
polynomials
satisfy the following differential  equation:

$$\Mhat' J_{\lambda,K}=0.\tag 8.3$$

\proclaim{Theorem 8.1} The space of solutions of the equation
$\Mhat' f=0$ in $A^{\prime\What}_K$ is finite-dimensional, and the
basis in the  space of solutions is given by the normalized Jack's polynomials
$J_{\lambda,K}$ (basis over $\C$, not over $A_0^{\prime \What}$!).
\endproclaim

\demo\nofrills{Proof}{} of this theorem is quite standard and is based
on the fact that each solution is uniquely determined by its highest
term.\enddemo

So far, all our constructions were purely algebraical; everything was
considered as formal power series in $q$. However, in order to study
modular properties we will need analytical approach. So, let us consider
every $e^\lhat\in\C[\Pp]$ as a function on the domain
$Y=\h\times\C\times\H$, where $\H$ is the upper half-plane:
$\H=\{\tau\in\C|\text{ Im }\tau >0\}$, by the following rule:
if $\lhat=\lambda+a\delta+K\eps$ then put
$e^\lhat(h,u,\tau)=e^{2\pi\i[\<\lambda,h\>+Ku-a\tau]}$. Note that this
agrees with our previous convention $e^{-\delta}=q$ if one lets
$q=e^{2\pi\i\tau}$.

Of course, we can't extend this rule to the completion $A$. However,
it turns out that we can extend it to certain elements of $A'$, namely
to the generalized characters:

\proclaim{Theorem 8.2}
For every $\lhat\in \Phat^+$, the Jack polynomial
$J_\lhat$ defined in Section~6, can be considered as an
analytical function on $Y$ by the above rule (i.e., the corresponding
series converges uniformly on compact sets in $Y$). The same is true for
$\hat\delta'$. \endproclaim

\demo{Proof}
The fastest way to prove this theorem is to use the defining differential
equation for $J_\lhat$. Indeed, due to Theorem~5.2 we can write
$J_\lhat=\sum_{\l\in P^+_{K}} f_\l(q)m_{\l, K}$
for some $f_l\in A'_0$. Substituting it in the defining differential
equation for $J_\lhat$ and
using the fact that $m_\lhat$ are eigenfunctions of $\hat\Delta$, we
get a system of ordinary differential equations for $f_\l$. It is easy
to check that the coefficients of these equations will be analytical
functions of $\tau$. Thus, we get that $f_\l$ will be analytical
functions of $\tau$.\enddemo

So, we can consider the generalized characters as functions on $Y$.
Note that this is equivalent to writing:

$$\chi_\lhat(h,u,\tau) =e^{2\pi\i K' u}\Tr|_{L_{\mu, K'}}
(\Phi q^{L_0-\frac{c}{24}} e^{2\pi\i h}),$$
where $K'=K+(k-1)h\v, \mu=\lambda+(k-1)\rho$, and
$L_0=-d+\text{const}$ where  the constant is chosen so that on the
highest weight vector, $L_0v_{\mu, K'}=
\frac{(\mu, \mu+2\rho)}{2(K'+h\v)} v_{\mu, K'}$,
and $c$ is
the Virasoro central charge: $c=\frac{K'\text{dim }\g}{K'+h\v}$.
Similar expressions appear in the Wess-Zumino-Witten
model of conformal field theory.

We can define the action of the affine Weyl group $\What$ on $Y$ so
that $e^{w\lhat}(h,u,\tau)=e^\lhat(w^{-1}(h,u,\tau))$. One easily
checks that when restricted to the finite Weyl group $W$ this action
coincides with the usual action of $W$ on $\h$ (leaving $u,\tau$
invariant), and the action of $\a\v\in Q\v$ is given by

$$\a\v(h,u,\tau)=(h-\a\v\tau, u+\frac 1 2
(\a\v,\a\v)\tau-\<\a\v,h\>,\tau).$$

This implies

\proclaim{Proposition 8.3} Let $f\in A_K^{\prime}$ be  such that
it gives an analytic function on $Y$. Then $f\in A_K^{\prime \What}$
iff the corresponding function on $Y$  satisfies the
following conditions:
\roster
\item"(a)"$f(h+\a\v,u,\tau)=f(h,u,\tau)$ for every $\a\v\in Q\v$.

\item"(b)"$f(h+\tau\a\v,u+\frac 1 2(\a\v,\a\v)\tau +\<\a\v,h\>,\tau)=
f(h,u,\tau)$ for every $\a\v\in Q\v$.

\item"(c)" $f(h, u+a,\tau)=e^{2\pi\i Ka}f(h,u,\tau)$ for every $a\in
\C$.

\item"(d)" $f(wh, u,\tau)=f(h,u,\tau)$ for every $w\in W$.
\endroster
\endproclaim

Note that conditions (a)--(c) can be rewritten as follows:
$f(h,u,\tau)=e^{2\pi\i Ku}g(h,\tau)$, and $g$ satisfies
$$\aligned
g(h+\a\v,\tau)&=g(h,\tau)\\
g(h+\a\v\tau,\tau)&=e^{-2\pi \i K(1/2 (\a\v,\a\v)\tau +
\<\a\v,h\>)}g(h,\tau),
\endaligned \tag 8.4$$
which is the standard definition of theta-functions. For this reason,
we will call analytical functions on $Y$ satisfying conditions
(a)--(c) above {\it theta-functions of level $K$} (cf. \cite{Lo}).

In a similar way, we can consider the differential operators $\hat L,
\Mhat, \Mhat'$ as usual differential operators on $Y$ with
analytical coefficients. For example, $\hat L$ can be rewritten as
follows:

$$\hat L= \frac{1}{4\pi^2} \biggl(-\sum \d_{x_i}^2+2\d_u\d_\tau\biggr)
 -k(k-1)\sum_{\a\in
R^+} (\a,\a)\varphi(\<\a,h\>, \tau),\tag 8.5$$
where, as before, $x_i$ is an orthonormal basis in $\h$ and

$$\varphi(x,\tau) = \sum_{m\in\Z} \frac{q^m e^{2\pi \i x} }{(1-q^m
e^{2\pi\i x})^2}=\frac{1}{4\pi^2} \d_x^2 \text{log }
\theta_1(x,\tau),$$
where we use the convention $q=e^{2\pi\i \tau}$ and

$$\theta_1(x,\tau)
= \i q^{1/8} (e^{\pi\i x}-e^{-\pi\i x})\prod_{n=1}^\infty (1-q^ne^{2\pi\i
x})(1-q^ne^{-2\pi\i x})(1-q^n).$$

 Note that the elliptic properties of
$\theta_1$ imply that
$\varphi(x+1,\tau)=\varphi(x+\tau,\tau)=\varphi(x,\tau)$; in fact,
$\varphi(x,\tau)=-\frac{1}{4\pi^2}\wp(x,\tau) +c(\tau)$, where
$\wp(x,\tau)$ is the Weierstrass function with periods $1,\tau$.
Thus,  the
operator $\hat L$ is well-defined on the torus $\h/(Q\v+\tau Q\v)$.
In a similar manner, one can consider $\Mhat, \Mhat'$ as
operators on $Y$.

\proclaim{Theorem 8.4}

\roster \item
For any $K\in\Z_+$, the Jack's polynomials
$\{J_{\lambda+K\eps}\}_{\lambda\in P_K^+}$ form a basis of the
space of analytical functions on $Y$ satisfying
conditions \rom{(a)--(d)} above over holomorphic functions of $\tau$.

\item For any $K\in\Z_+$, the normalized   Jack's polynomials
$\{J_{\lambda,K}\}_{\lambda\in P_K^+}$
form a basis of the
space of analytical functions on $Y$ satisfying
conditions \rom{(a)--(d)} above  and the condition $\Mhat'f=0$ over
$\C$. \endroster\endproclaim

\subheading{9. Modular invariance}

Recall that the modular group $\Gamma=SL_2(\Z)$ is generated
by the elements
$$S=\pmatrix 0&-1\\ 1 &0\endpmatrix, T=\pmatrix 1&1\\ 0
&1\endpmatrix$$
satisfying the defining relations $(ST)^3=S^2, S^2T=TS^2, S^4=1$.
This group acts in a natural way on $Y$ as follows:

$$\pmatrix a&b\\c&d\endpmatrix (h,u,\tau)=\left(\frac{h}{c\tau+d},
u-\frac{c(h,h)}{2(c\tau+d)}, \frac{a\tau+b}{c\tau+d}\right)\tag 9.1$$

In particular,

$$\align
T(h,u,\tau)&=(h, u,\tau+1)\\
S(h,u,\tau)&=\left(\frac h \tau , u-\frac{(h,h)}{2\tau} ,
-\frac{1}{\tau}\right)\endalign$$

Also, for any $j\in \C$ we will  define a right action of $\Gamma$
on functions on $Y$ as follows: if $\alpha=\pmatrix
a&b\\c&d\endpmatrix$ then let

$$(f[\alpha]_j)(h,u,\tau)=(c\tau+d)^{-j} f(\alpha(h,u,\tau)).$$

In fact, this is a projective action, which is related to the
ambiguity in the choice of $(c\tau+d)^{-j}$ for non-integer $j$; to
make it a true  action one must consider a central extension of
$SL_2(\Z)$;  we are not going into details here. We will call this
action ``an action of weight~$j$''.

Our main goal will be to find the behaviour of the (normalized) affine
Jack's polynomials under modular transformations. The first result in
this direction is
\proclaim{Theorem 9.1} Fix $k, \varkappa \in Z_+$.
 Then the space of all solutions
of the equation $\hat L f=0$ in the space of
theta-functions of level $\varkappa$ on $Y$ (i.e., functions $f$ satisfying
the conditions \rom{(a)--(c)} of Proposition \rom{8.3}) is invariant under
the action  of $\Gamma$ of
weight $j=\frac r 2\left(1+\frac{k(k-1)h\v}{\varkappa}\right)$. \endproclaim

\demo{Proof} It is easy to see that the operator $\hat L$ is invariant
under the action of $T\in \Gamma$. Thus, to prove the theorem, it
suffices to prove the following formula:

$$\hat L (f[S]_j)=\tau^{-2} ((\hat L g)[S]_j)-
\frac{1}{2\pi\i\tau}(\varkappa(r-2j)+k(k-1)rh\v) f[S]_j.$$

This is based on formula (8.5) for $\hat L$. Indeed, using modular
properties of the theta-function (see, for example, \cite{Mu}), we can
show that $\varphi(\frac x\tau, -\frac 1\tau)= \tau^2\varphi(x,
\tau)+\frac{\i}{2\pi}\tau$.

Also, it is not too difficult to check that

$$\hat\Delta \left( f[S]_j\right) = \tau ^{-2}( (\hat\Delta f)[S]_j)
- \frac{\varkappa}{2\pi\i\tau}(r-2j)(f[S]_j),$$
where $\hat\Delta= 1/4\pi^2(-\sum\d_{x_i}^2 +2\d_u\d_\tau)$.

Since $\d_u f=2\pi\i \varkappa f$ and $\sum_{\a\in R^+}(\a,\a)=rh\v$, we get
the desired formula. \enddemo

Next, we will need the following well-known fact (see \cite{K}):
$$\hat\delta'[\a]_{r/2}=l(\a)\hat\delta',$$
where the function (not a character)
$l:\Gamma\to \C^\times$ is such that $l^{24}=1$.

This gives us the following theorem:
\proclaim{Theorem 9.2} For a fixed $k\in \Z_+$, the linear span of the
normalized Jack's polynomials $J_{\lambda,K}, \lambda\in P^+_K$ is
invariant under the action of $\Gamma$ of weight
$j=-\frac{K(k-1)r}{2(K+kh\v)}$\endproclaim

\demo{Proof} This is a corollary of Theorem~8.4, definition of
$\Mhat'$  and the previous theorem.
\enddemo

Thus, for every $K\in\Z_+$ we have a projective representation of
$\Gamma$ in the finite-dimensional linear space
$V_K=\bigoplus\limits_{\lambda\in P^+_K} \C J_{\lambda,K}$; in fact,
we have a family of representations of
$\Gamma$ in the same space $V_K$,
which are obtained for different values of $k$.

In general, these representations seem
to be very interesting. First of all, note that it follows from the
formula for the highest term of $J_{\l,K}$ that the eigenvalues of $T$ in
such a  representation are roots of unity
of degree $N=M(K+kh\v)$, where $M$ is the smallest positive integer such
that $(\lambda, \mu)\in \frac{1}{M} \Z, (\lambda, \lambda)\in
\frac{2}{M}\Z$ for any $\lambda, \mu\in P$.
Thus, for fixed $K$ and  large enough $k$  the order of
$T$ tends to $\infty$.
It is known that for ordinary characters, the representation of the
modular group (considered as
mapping $\Gamma\to PGL(V_K)$) is trivial on the  congruence subgroup
$\Gamma(N)$ and thus is in fact a representation of the finite group
$\Gamma/\Gamma(N)\simeq SL_2(\Z/N\Z)$. For generalized characters
 it is not so. The best we can say is the following trivial
proposition:

\proclaim{Proposition} The
representation of $\Gamma$ in $V_K$ is trivial on the normal subgroup $T(N)$,
which by definition is the smallest normal subgroup containing
$T^{N}, N=M(K+kh\v)$.\endproclaim

\example{Example} Let $\g=E_8$. Then for any fixed $K>1$ coprime with
$h\v=30$ there is an infinite number of values of $k$ for which the
image of $\Gamma$ in $PGL(V_K)$ is infinite. \endexample

\demo{Proof}
 This is based on the following theorem, due to Jordan (cf.
\cite{CR, \S36}): for every
fixed $n$ there is a constant $C(n)$ such that any finite subgroup in
$PGL(n,\C)$ has a commutative normal subgroup of index not exceeding
$C(n)$. Now, fix $K$; take the constant $C=C(|P^+_K|)$ and choose $k$
such that all the prime factors of $N=K+kh\v$ are larger
than $C$ (recall that $M=1$ for $E_8$).
Assume that the image of $\Gamma$ in $PGL(V_K)$ is finite;
then it has a commutative
normal subgroup $A$ of index $\le C$. Since $T^N=1$,
the order of $T$ is relatively prime with the index of $A$. Thus, image of
$T$ is in $A$. But the same is true for $T'=S^{-1}TS=\pmatrix
1&0\\1&1\endpmatrix$, since it is conjugate with $T$. On the other
hand, $T$ and $T'$ generate $\Gamma$, and thus the image of $\Gamma$
is contained in $A$ and thus is commutative. But it is known that
$|\Gamma/[\Gamma,\Gamma]|=12$, and thus $\Gamma$ cannot have a
commutative quotient of order greater than $12$.
\qed\enddemo

\remark{Remark} It seems plausible that for any $\g$ and
fixed sufficiently large  $K$
the image is in fact infinite
for all sufficiently large $k$. \endremark

It is known that the quotient  $\Gamma(N)/T(N)\simeq\pi_1(\Sigma_N)$, where
$\Sigma_N$ is the modular curve: $\Sigma_N=\overline{\H/\Gamma(N)}$.
Thus representations of $\Gamma(N)/T(N)$
classify the flat connections in vector bundles
over $\Sigma_N$, and representations of
$\Gamma/T(N)$ classify the flat connections that are invariant with
respect to the natural action of $SL_2(\Z/N\Z)$ on $\Sigma_N$.
It seems
interesting to interpret our representation from this point of view.
Note that the usual characters from this point of view are trivial, so
this phenomenon is specific for $k>1$.

\subheading{10. Unitarity and relation with conformal field theory}

This section is devoted to the discussion of the following conjecture.

\proclaim{Conjecture 10.1} For every $K,k\in \Z_+$ there exist positive
real numbers $d_\lambda, \lambda\in P^+_K$ such that the above
defined projective
action of $\Gamma$ in the space $V_K$ is unitary with respect to the
hermitian form in $V_K$ defined by $(\hat J_{\lambda, K},\hat J_{\mu,
K})= d_\lambda \delta_{\lambda, \mu}$. \endproclaim

\demo{Example} For $k=1$ this conjecture holds with $d_\lambda=1$; the
proof is based on the Weyl-Kac formula for the characters
(see\cite{K}).\enddemo

\proclaim{Corollary} Conjecture \rom{10.1} implies that the
eigenvalues of the action in $V_K$ of any $x\in
\Gamma$ lie have unit norm. \endproclaim

\demo\nofrills{}
Note that this makes sense: though it is a projective representation,
the corresponding cocycle takes values in the
unit circle, and
thus does not change the notion of unitarity.

This conjecture is motivated by the modular invariance in
conformal field theory (CFT). We briefly outline the relation here; for
detailed exposition, see, for example, \cite{MS}.

The Wess-Zumino-Witten model of conformal field theory is based on the
integrable highest-weight representations of $\ghat$ of level $K$. The
space of physical states in this theory is the Hilbert space

$$H=\bigoplus_{\lambda\in P^+_K} L_{\lambda, K}\otimes L_{\lambda^*,K},$$
where the involution ${}^*: P^+\to P^+$ is defined by the condition
that the finite-dimensional modules over $\g$ are related by
$V_{\lambda^*}= \left(V_{\lambda}\right)^*$.

The essence of the CFT is construction of the so-called amplitudes.
We consider Riemann surfaces $\Sigma$ together with a finite number of
marked points and local parameters at these points, divided into two
subsets of ``incoming'' and ``outgoing'' points. To each marked point
we associate a copy of the space $H$  and define the
spaces $H_{in}, H_{out}$ which are just the tensor products of the spaces
$H$ over all the incoming (resp., outgoing) points. Then we must
construct the {\it amplitudes}, i.e.,
the  operators
$A(\Sigma):H_{in} \to H_{out}$, satisfying a number of axioms; the
most important of them is the sewing axiom.

In the WZW model these amplitudes are defined as follows: first, we
consider a slightly more general setting and consider Riemann surfaces
with marked points (divided into ``incoming'' and ``outgoing''), local
parameters at these points and a choice of $\lambda\in P^+$ for every
marked point. Then we construct a map

$$A_{\lambda_1, \ldots; \mu_1, \ldots}:
L_{\lambda_1, K}\otimes L_{\lambda_2, K}\otimes \dots\to
	W_{\lambda_1, \ldots, \mu_1, \ldots} \otimes
L_{\mu_1,K}\otimes\dots,\tag 10.1$$
where the tensor products are taken over  all incoming (resp.,
outgoing) points and $W$ is some finite-dimensional space, depending
on $\Sigma, \lambda_i, \mu_j$, called the space of conformal blocks.
Now we can construct the global amplitude $A$ as follows:

$$A=\bigoplus_{\lambda, \mu} (A_{\lambda_i, \mu_j}, A_{\lambda_i^*,
\mu_j^*}),\tag 10.2$$
and $(\cdot, \cdot)$ stands for a Hermitian form

$$W_{\lambda_i,
\mu_j}\otimes W_{\lambda^*_i, \mu^*_j}\to \C.\tag 10.3$$

Thus, to define the amplitudes we must define the spaces of conformal
blocks along with the mappings (10.1) and pairing (10.3).
The sewing axiom along with some other axioms of CFT says that all  of
these is defined uniquely as soon as it is defined for the sphere with
three punctures at $0, z_0, \infty$ and local parameters $z, z-z_0,
1/z$. In this case, it is possible to define the space of conformal
blocks explicitly. Namely, if $0$ is the incoming point with the
weight $\lambda$ assigned to it and $\infty, z_0$ are outgoing points
with the assigned weights $\mu, \nu$ respectively then one can define
the corresponding space of conformal blocks as the space of all vertex
operators, i.e., operators

$$\Phi: L_{\lambda, K}\to L_{\mu, K}\otimes L_{\nu, K}$$
satisfying certain commutation relations (cf. \cite{MS,TK}). It is
known that such an operator is uniquely defined by its restriction to
the highest level (with respect to the $d$-grading) of the module
$L_{\nu, K}$, and this restriction must be the intertwiner for $\ghat$
if we consider the highest level of $L_{\nu,K}$ as the evaluation
representation of $\ghat$; thus, it is the same  intertwining operator which
we considered in Section~7. However, the question of defining the
scalar product on the space of conformal blocks is much more subtle
(cf. \cite{FGK}). Nonetheless, it is generally believed that the
following is true.

\proclaim{Conjecture: consistency  of WZW model} For a suitably defined inner
product on the space of conformal blocks, the resulting conformal
field theory is well-defined, i.e., the amplitudes \rom{(10.2)}
are uniquely determined by the complex structure on $\Sigma$ and
do not depend on the choice of obtaining $\Sigma$ by gluing from
three-punctured spheres, caps and cylinders. \endproclaim

This is a very strong condition. It was shown in \cite{MS} that it
suffices to check consistency for the sphere and the torus. Moreover,
for the sphere it is proved in the framework of vertex operator algebras.
However, to the best of our knowledge, no satisfactory proof is known
for consistency of WZW model on the torus; yet, some physical
arguments (such as path integration) suggest that
this  is indeed true.

Returning to affine Jack's polynomials, we can say that the unitarity
conjecture 10.1 can be deduced from the conjecture on the consistency
of the WZW model formulated above. Indeed, it can be easily checked
that the unitarity of the action of $\Gamma$ is equivalent to modular
invariance of the function

$$F(\tau)=\sum_{\lambda\in P^+_K} d_\lambda J_{\lambda, K}(0, 0, \tau)
\overline{J_{\lambda, K}}(0,0,\tau). $$

This function is nothing but a certain  component of the
one-point correlation function on a torus for the Wess-Zumino-Witten
conformal field theory, based on integrable representations of $\ghat$
of level $K+(k-1)h\v$, and thus, its modular invariance follows from
consistency of the Wess-Zumino-Witten model which in
particular implies that the correlation function does not
depend on the choice of
representation of  the torus in the form $T=\C/(\Z+\tau\Z)$, i.e., is
modular invariant.

\enddemo

\subheading{11. Quantum affine algebras and affine analogue of the
Macdonald's polynomials}
In this section we briefly outline how to define the generalized
characters for quantum affine algebra. Let $\U$ be the quantum affine
algebra, i.e., the quantization of the universal enveloping algebra of
$\widehat{\frak{sl_n}}$ (see \cite{Dr, J1, J2}). We use $p$ for the
quantization parameter to avoid confusion with $q$ used in previous
section for denoting the modular parameter of the torus. We can define
for this algebra the notion of Verma module, irreducible
highest-weight module etc.  in the same manner as for usual affine
algebra. As before, the modules $L_\lhat, \lhat\in \Phat^+$ are called
integrable modules. Note that unlike the classical case, there is no
natural action of (central extension of)
$\What$ in integrable modules (though, of course,
there is an action of $\What$ on the set of weights of $L_\lhat$).
Also, we can define the notion of evaluation representation $V(z)$,
though it is much less obvious; it is based on the existence of the
evaluation homomorphism $\U\to U_p\frak{sl}_n$ (see \cite{J2}).

Similar to Section~7, define intertwiners
$$\Phi^p_\lhat \colon L_{\hat\mu}\to \widehat{L_{\hat\mu}}\otimes V(z),$$
where $V$ is the deformation of representation of $\frak{sl}_n$ in
$S^{(k-1)n} \C ^n$, and $\hat\mu=\lhat+(k-1)\rhat, \lhat\in \Phat^+$.
Also, define the corresponding traces:

$$\chi_\lhat=\sum_{\hat\nu\in \Phat}e^{\hat\nu}\Tr_{L_{\hat\mu}[\hat\nu]}
\Phi^p_\lhat.\tag 11.1$$

Most of the theory developed for these traces in Section~7 can be
generalized to the quantum case with some changes. However, there are two
major distinctions. First, these traces do not have any natural
$\What$-symmetry; however, we never used the $\What$-symmetry of the
traces in Section~7; to prove the symmetry of the ratio
$\chi_\lhat/\chi_0$ we only used $\What$-invariance of usual characters,
which is still valid for quantum case. Another distinction is that we do
not have any differential equation for these traces. In principle, one can
say that the quantized traces satisfy some difference equation, which is
an analogue of (7.3), but  we do not know the explicit
expression. However, we still have the following propositions.

\proclaim{Theorem 11.1}
$$\chi^p_0=f(p,q)e^{(k-1)\rhat}\prod_{j=1}^{k-1}\prod_{\ahat\in \Rhat^+}
(1-p^{2j}e^{-\ahat}),\tag 11.2$$
where $f$ is some formal power series in $q=e^{-\delta}$ whose
coefficients  are rational functions in $p$ and highest
term is $1$. \endproclaim

Proof of this theorem is quite similar to that of Theorem~7.2 and can be
obtained by conjunction of arguments in the proofs of Theorem~7.2 and
Proposition~4.1 in
\cite{EK2}; however,
in the quantum case we do not have the differential equation
for traces, and thus are unable to determine the factor $f(p,q)$.

\proclaim{Theorem 11.2}
The ratio $\hat J^p_\lhat=\chi^p_\lhat/\chi^p_0$ is a symmetric Laurent
polynomial: $J^p_\lhat\in A^\What$, and has highest term $e^\lhat$.
\endproclaim

Again, the proof is a literal repetition of the proof of Theorem~7.3.

\definition{Definition} The polynomials
$\hat J^p_\lhat=\chi^p_\lhat/\chi^p_0$ are called affine Macdonald's
polynomials. \enddefinition

This definiton is motivated by the fact that the same construction for
the finite-dimensional case (i.e., for representations of $U_p\g$)
gives the usual Macdonald's polynomials.
We believe that the above definition givs   the right affine  analogue
of Macdonald's polynomials. However, at this moment we are unable to
define them as orthogonal with respect to a certain inner product or as
eigenfunctions of an explicitly presented  difference operator.

Note that among the identities satisfied by Macdonald's polynomials
(see \cite{M2}) there is one that can be easily generalized to the
affine case: this is the
so-called special value identity, which for simply-laced case looks as
follows: if one defines the evaluation map $\pi: e^\lambda\to
p^{2(\lambda, k\rho)}$ then
$$\pi (P_\lambda)=p^{-2(\lambda, k\rho)}
\prod_{\a\in R^+} \prod_{i=0}^{k-1}
\frac{1-p^{2(\a, \lambda+k\rho)+2i}}{1-p^{2(\a,k\rho)+2i}}.$$

This leads us to the following conjecture:
\proclaim{Conjecture 11.3} The affine Macdonald's polynomials $\hat
J^p_\lhat$ satisfy the following  identity:

$$\pi(\hat J^p_\lhat)= p^{-2(\lhat, k\rhat)}
\prod _{\ahat\in \Rhat^+} \prod_{i=0}^{k-1}
\frac{1-p^{2(\ahat, \lhat+k\rhat)+2i}}{1-p^{2(\ahat, k\rhat)+2i}},$$
where, as before, the evaluation $\pi$ is defined by $\pi(e^\lhat)=
p^{2(\lhat, k\rhat)}$.
\endproclaim

\demo{Example} For $k=1$ this formula can be easily proved using the
Weyl-Kac character formula and the denominator identity for affine Lie
algberas. \enddemo

Finally, we briefly mention  how to generalize the construction  of the affine
Macdonald's polynomials to generic $k$ (not necessarily positive integer).
This is done quite similar to the finite-dimensional case (see
\cite{EK2}). First of all, it is easy to show that the affine
Macdonald's polynomials have the form $P=\sum_{\lhat} f_\lhat(p,p^k)
e^\lhat$, where $f_\lhat(p,t)$ is a rational function of two variables; it
does not depend on $k$. Thus, it is very easy to continue such a
polynomial to arbitrary value of $k$ (provided that $f$ does not have
a pole at this point). It is convenient to consider $k$ as a formal
variable, i.e., consider $p$ and $p^k$ as algebraically independent;
we will use this convention in the remaining part of the paper.

Consider the intertwining operator

$$\tilde\Phi_\lhat: M_{\lhat+(k-1)\rhat} \to M_{\lhat+(k-1)\rhat}\otimes U_k,$$
where $M_{\lhat+(k-1)\rhat}$ is Verma module and $U_k$ is the
evaluation representation of $\U$ obtained from the standard  action of
$U_p\frak{sl}_n$ in the functions of the form
$(x_1\ldots x_n)^{k-1}p(x)$, $p(x)$ being a Laurent polynomial in
$x_i$ of total degree zero (see \cite{EK2}). Since $k$ is a formal variable,
$M_{\lhat+(k-1)\rhat}$ is irreducible and thus such an intertwiner
exists and is unique up to a constant.
Similarly to Section~7, define the traces

$$\tilde\chi_\lhat=\sum_{\hat\mu}e^{\hat\mu}
	\Tr_{M_{\lhat+(k-1)\rhat}[\hat\mu]}(\tilde\Phi_\lhat)\tag 11.3$$

\proclaim{Theorem 11.4} \roster \runinitem
The ratio $\tilde\chi_\lhat/\tilde\chi_0$ is
the affine Macdonald's polynomial $\hat J^p_\lhat$.

\item
$$\tilde\chi_0=f(p,q) e^{(k-1)\rhat}\prod_{\ahat\in\Rhat^+}\prod_{i=1}^\infty
\frac{1-p^{2i}e^\ahat}{1-p^{2(k-1)+2i}e^\ahat}.$$
\endroster
\endproclaim

\demo{Proof} The proof is quite similar to the finite-dimensional case
(see \cite{EK2, Section~5}) and is based on the fact that for fixed
$\lhat, \hat\mu$ the weight subspace
$L_{\lhat+(k-1)\rhat}[\lhat+(k-1)\rhat-\hat\mu]$ (here $k$ is a
positive integer) is for $k$ large enough isomorphic to the weight subspace
$M_{\lhat+(k-1)\rhat}[\lhat+(k-1)\rhat-\hat\mu]$, where $k$ is a formal
variable, and the restriction to this subspace
of the intertwiner $\Phi$, defined in
Section~7 for $k\in\Z_+$ coincides for $k$ large enough  with the
specialization of the intertwining operator $\tilde\Phi$ defined above
for the case when $k$ is a formal variable. On the other hand, if we
have two rational functions of two variables $f(p,t), g(p,t)$ such
that $f(p,p^k)=g(p,p^k)$ for all integer $k>>0$ then $f=g$ as functions of two
variables. We refer the reader to \cite{EK2} for the details. \enddemo

\subheading{12. Summary}

Let us summarize what has been done so far. We defined the affine analogue of
Jack's and Macdonald's polynomials (for the root system $\hat
A_{n-1}$); we also checked that the affine Jack's polynomials are
eigenfunctions of
the affine analogue of Sutherland operator. If we now turn back to the
finite-dimensional case, we see that in the affine case there are
still missing pieces of structure. First of all, we have not defined the inner
product which would make these polynomials orthogonal. This is a
deep problem; the trivial generalization of the finite-dimensional
inner product  is not convergent. However, there are good
reasons to believe that one can regularize this inner product so to
make sense of it. Indeed, if we write the affine analogue of
Macdonald's inner product identities, which give an explicit
expression for $\<P_\lambda, P_\lambda\>_k$ in terms of the root
system, then the product in the right-hand side has a pole as $p\to 1$
(recall that $p$ is the parameter of the quantization). However, for
fixed $K,k$ the order of the pole is the same for all $\lambda$. This
suggests that one can renormalize the trivial definition of the inner
product (for a fixed
level) and then prove  the affine version of Macdonald's inner product
identities. In this case the right-hand side of these identities is
expressed in terms of $\Gamma$-function at  rational points.
Moreover, there are certain arguments suggesting that
this renormalized inner product should coincide
with the inner product discussed in Section~10, i.e., the inner
product known in CFT. Remarkably, in those few cases when the inner
product on the space of conformal blocks of CFT is written down
explicitly (see \cite{MS, Appendix D}), the answer is also written as
a ratio of  gamma-functions of rational argument.
This will be discussed in our future papers.

Finally, in the finite-dimensional case we had a large family of
commuting differential (for $q\ne 1$, difference) operators. In the
affine case similar operators can be defined (see \cite{E1}). However,
these operators are not difference operators but rather infinite sums
of difference operators.

If we pass to the  critical level
($K+kh\v=0$) then the (local completion of) the universal enveloping algebra
has a center. In this case the term containing $q\frac{\d}{\d q}$ in
the expression for the affine Sutherland operator drops, and we get
the elliptic Calogero-Sutherland operator, which can be included in
a family of commuting differential operators (see \cite{OP, Ch}),
 which can be obtained from the elements of the center
(see \cite{E}).  This
suggests that the
asymptotics of affine Jack's polynomials in the limit $K$ fixed,
$K+kh\v\to 0$  should be very interesting. These asymptotics can be
calculated by writing explicit integral formulas for the traces of
intertwiners (similar to \cite{BF, EK3}) and finding the asymptotics
by the saddle-point method. For the case $\g=\frak{sl}_2$ it was done
in \cite{EK3}; general case will be discussed
in our future papers.

\enddocument
\end